\documentclass[preprint]{aastex}

\newcommand{\lae}{\mathrel{<\kern-1.0em\lower0.9ex\hbox{$\sim$}}}
\newcommand{\gae}{\mathrel{>\kern-1.0em\lower0.9ex\hbox{$\sim$}}}
\newcommand{\kms}{\mbox{ km s}^{-1}}
\newcommand{\kpc}{\mbox{ kpc}}
\newcommand{\cm}{\mbox{ cm}}
\newcommand{\yr}{\mbox{ yr}}
\newcommand{\area}{\mbox{ cm}^{2}}
\newcommand{\K}{\mbox{ K}}
\newcommand{\s}{\mbox{ s}}
\newcommand{\pcc}{\mbox{ cm}^{-3}}
\newcommand{\flux}{\mbox{ ergs cm}^{-2}\mbox{ s}^{-1}}
\newcommand{\La}{Ly$\alpha$}
\newcommand{\NV}{N\,{\sc v} $\lambda\lambda 1239, 1243$}
\newcommand{\SiIV}{Si\,{\sc iv} $\lambda\lambda 1394, 1403$}
\newcommand{\OIV}{O\,{\sc iv}] $\lambda 1400$}
\newcommand{\NIV}{N\,{\sc iv}] $\lambda 1487$}
\newcommand{\CIV}{C\,{\sc iv} $\lambda\lambda 1548, 1551$}
\newcommand{\HeII}{He\,{\sc ii} $\lambda 1640$}
\newcommand{\OI}{[O\,{\sc i}] $\lambda\lambda 6300, 6364$}
\newcommand{\NII}{[N\,{\sc ii}] $\lambda\lambda 6548, 6584$}
\newcommand{\Ha}{H$\alpha$}
\newcommand{\HeI}{He\,{\sc i} $\lambda 6678$}
\newcommand{\SII}{[S\,{\sc ii}] $\lambda\lambda 6717, 6731$}

\begin{document}

\title{HST Spectroscopy of Spot 1 on the Circumstellar Ring of SN
1987A}

\author{Eli Michael\altaffilmark{1}, 
Richard McCray\altaffilmark{1}, 
C. S. J. Pun\altaffilmark{2}, 
Peter Garnavich\altaffilmark{3}, 
Peter Challis\altaffilmark{3}, 
Robert P. Kirshner\altaffilmark{3}, 
John Raymond\altaffilmark{3}, 
Kazimierz Borkowski\altaffilmark{4}, 
Roger Chevalier\altaffilmark{5}, 
Alexei V. Filippenko\altaffilmark{6}, 
Claes Fransson\altaffilmark{7}, 
Peter Lundqvist\altaffilmark{7},
Nino Panagia\altaffilmark{8}, 
M. M. Phillips\altaffilmark{9}, 
George Sonneborn\altaffilmark{2},
Nicholas B. Suntzeff\altaffilmark{10}, 
Lifan Wang\altaffilmark{11},
and J. Craig Wheeler\altaffilmark{11}}

\altaffiltext{1}{JILA, University of Colorado, Boulder, CO 80309-0440;
michaele@colorado.edu}
\altaffiltext{2}{Laboratory for Astronomy and Space Physics, Code 681,
NASA-GSFC, Greenbelt, MD 20771}
\altaffiltext{3}{Harvard-Smithsonian Center for Astrophysics, 60
Garden St, Cambridge, MA 02138}
\altaffiltext{4}{Dept. of Physics, North Carolina State University,
Raleigh, NC 27695}
\altaffiltext{5}{Dept. of Astronomy, University of Virginia, P.O. Box
3818, Charlottesville, VA 22903-0818}
\altaffiltext{6}{Dept. of Astronomy, University of California,
Berkeley, CA 94720-3411} 
\altaffiltext{7}{Stockholm Observatory, SE-133 36 Saltsj\"obaden,
Sweden} 
\altaffiltext{8}{STScI, 3700 San Martin Drive, Baltimore, MD 21218; on
assignment from the Space Science Department of ESA.}
\altaffiltext{9}{Las Campanas Observatory, Carnegie Observatories,
Casilla 601, La Serena, Chile}
\altaffiltext{10}{CTIO, NOAO, Casilla 603, La Serena, Chile}
\altaffiltext{11}{Dept. of Astronomy, University of Texas, Austin, TX 78712}

\begin{abstract}
We present ultraviolet and optical spectra of the first bright spot
(PA = $29^\circ$) on Supernova 1987A's equatorial circumstellar ring
taken with the Space Telescope Imaging Spectrograph.  We interpret
this spot as the emission produced by radiative shocks that occur
where the supernova blast wave strikes an inward protrusion of the
ring.  The observed line widths and intensity ratios indicate the
presence of radiative shocks with velocities ranging from $100$ to
$250 \kms$ entering dense ($\gae 10^4 \pcc$) gas. These observations,
and future observations of the development of the spectra and line
profiles, provide a unique opportunity to study the hydrodynamics of
radiative shocks.
\end{abstract}


\section{INTRODUCTION}

After the discovery of the circumstellar ring around Supernova
(SN) 1987A, several authors pointed out that the supernova blast wave
should strike the ring some 10 -- 20 years after the supernova
explosion \citep{LM91,LMS94,CD95,BBM97}. The estimated time of
first contact was uncertain, mainly because it depended on the unknown
density of circumstellar gas between the supernova and the ring.

This event is now underway.  In 1995 an unresolved spot at position
angle $\sim 29^\circ$ (Spot 1) on the near (North) side of the ring
began to brighten \citep{GKC97}. This spot continues to brighten, and
within the last year several new spots have appeared
\citep{LC00,GKC00,Lawr00}. A shock interpretation for these spots was
suggested by Space Telescope Imaging Spectrograph (STIS) observations
of the ring taken in 1997 \citep{Sonn98}.  The spectral image of the
ring in \Ha\ showed a streak at the position of Spot 1 indicating
blueshifted emission with velocities up to $\sim 250 \kms$.
Evidently, at the positions of the spots, the blast wave is striking
inward protrusions of the dense equatorial ring.

Here (\S 2) we report on early optical and ultraviolet STIS spectra of
SNR 1987A which show several broad lines associated with Spot 1. In \S
3 we present the radiative shock model for the emission from the
spots, and in \S 4 we discuss what we can infer from the spectrum and
its evolution.

\section{OBSERVATIONS}

The Supernova INtensive Study (SINS) collaboration obtained STIS
spectra of Spot 1 at ultraviolet and optical wavelengths in 1997
September and 1998 March, respectively.  The ultraviolet spectrum (not
shown) was taken on September 27, 1997 (3869.3 days since explosion)
with the G140L low dispersion grating (1150 -- 1736~\AA, $\Delta v
\approx 300 \kms$) and a $0\farcs5$ slit.  Five spectral images
totaling $11,200 \s$ were combined.  Emission lines from nearly
stationary gas in the ring are apparent at \NV, \OIV, \NIV, and
\HeII. Emission from Spot 1 is also visible in these lines, as well as
at \SiIV, [N\,{\sc iv}] $\lambda 1483$, and \CIV. The low-velocity
Si\,{\sc iv} and C\,{\sc iv} emission from the ring is blocked by
interstellar absorption \citep{Fran89}.  Due to geocoronal \La\
emission no information about \La\ from the spot was obtained.

The optical spectrum was taken on March 7, 1998 (4030.0 days since
explosion), with a $0\farcs2$ slit and the G750M grating (6295 --
6867~\AA, $\Delta v \approx 50 \kms$).  Three spectral images totaling
$8,056 \s$ were combined.  A section of the optical spectrum is shown
in Fig.~\ref{optspect}. The central horizontal streak (which extends
to $\sim \pm 3,000 \kms$) is \Ha\ emission from the radioactively
heated supernova debris.  Pairs of bright spots where the slit
intersects the stationary inner ring are evident in \NII\ and \Ha\
(not displayed but also present are \OI, \HeI, and \SII).  Three
fainter spots due to the outer rings are also visible in \Ha\ and
[N\,{\sc ii}].  Broad streaks of emission from Spot 1 appear on the
upper segments of the ring lines.  These streaks are inside the ring
emission indicating that Spot 1 lies at the inner edge of the ring.

\medskip
\begin{center}
\includegraphics[width=3.25in]{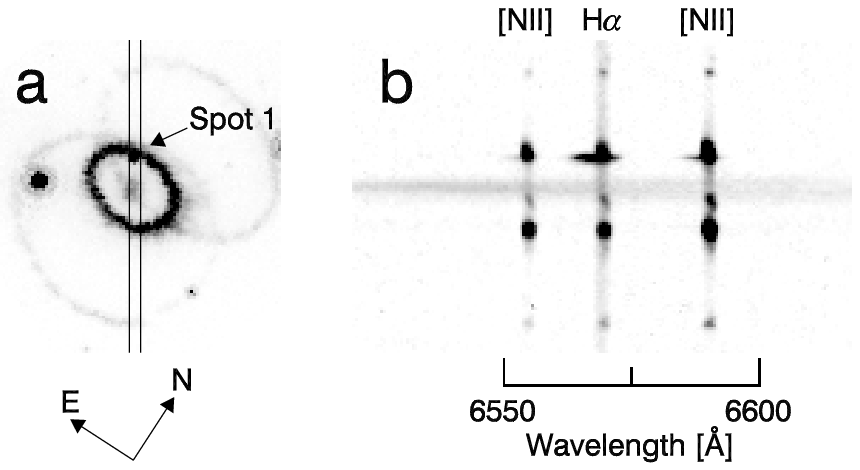}
\figcaption[f1.eps]{(a) Slit orientation on image of SN 1987A's
triple ring system. (b) Section of STIS G750M optical spectrum (the
intensity scale has been stretched to show the hot spot
emission).\label{optspect}}
\end{center}
\medskip

The spectra of Spot 1 are extracted from the STIS spectral images. For
the optical emission lines, the major source of contamination is the
emission from the nearly stationary ($v = 10.5 \pm 0.3 \kms$,
\citeauthor{CH99} \citeyear{CH99}) circumstellar ring filling the
$0\farcs2\:(\approx 100 \kms)$ slit.  The \Ha\ line profile is shown
in Figure~\ref{haprofile}.  The parts of the line profiles that lie
outside of the ring emission are fit well by Gaussians.  The velocity
shifts of the Gaussian fits are poorly constrained since the spot's
spatial position in the slit is uncertain. In the ultraviolet the ring
emission is much weaker and therefore its contribution to the
extracted Spot 1 spectrum is neglected.  The UV lines are poorly
resolved at the resolution of the G140L grating.  It is possible to
account for the detector line spread function in order to obtain
intrinsic widths for the UV lines \citep{Pun00}.  While this method
introduces errors in the determined width, we find that all the UV
lines have intrinsic widths $< 400 \kms$.

\medskip
\begin{center}
\includegraphics[width=3.25in]{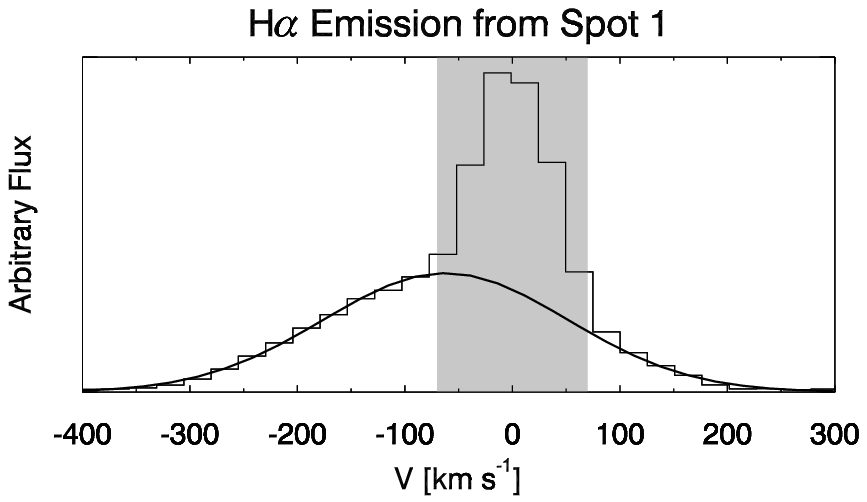}
\figcaption[f2.eps]{\Ha\ profile of Spot 1 extracted from the G750M
spectrum.  Low velocity emission (shaded region) from unshocked
stationary ring material defines the standard of rest.  A Gaussian fit
to the broad emission from shocked gas is shown.\label{haprofile}}
\end{center}
\medskip

All the lines from Spot 1 were fit with Gaussians in order to extract
their fluxes and full widths at half maximum (FWHM) (see Table 1).
The errors stated in Table 1 are $1\sigma$ statistical errors to the
fits. Close multiplets (\NV\ and \CIV) were fit simultaneously by
requiring a set wavelength separation, identical widths, and line
ratios dictated by their oscillator strengths.  For the N\,{\sc v}
doublet, the emission from Spot 1 overlapped with very broad \La\
emission ($\sim \pm 15,000 \kms$) from the reverse shock
\citep{Sonn98,Mich98}. This \La\ background was removed using a
quadratic fit and is the source of the large uncertainty in the \NV\
flux.  At the G140L resolution, the Si\,{\sc iv} $\lambda 1403$
emission of Spot 1 is mixed with that from the \OIV\ multiplet.  Using
the measured Si\,{\sc iv} $\lambda$1394 flux and assuming
$I(1394)/I(1403) = 2$, the Si\,{\sc iv} $\lambda 1403$ flux was
determined and removed from the \OIV\ multiplet.

\medskip
\begin{center}
\centerline{Emission Lines from Spot 1}
\begin{tabular}{lcc}
\tableline\tableline
Emission Line                   & FWHM$^a$      & Observed Flux$^b$     \\
\tableline
\NV                             & \nodata       & $2.5 \pm 0.9$         \\
\SiIV                           & \nodata       & $0.6 \pm 0.2$         \\ 
O\,{\sc iv}] $\lambda 1400$     & \nodata       & $1.3 \pm 0.4$         \\ 
$[$N\,{\sc iv}] $\lambda 1483$  & \nodata       & $1.4 \pm 0.4$         \\
\NIV                            & \nodata       & $1.5 \pm 0.4$         \\ 
\CIV                            & \nodata       & $2.9 \pm 0.8$         \\ 
\HeII                           & \nodata       & $3.5 \pm 0.9$         \\ 
\tableline\tableline
$[$O\,{\sc i}] $\lambda 6300$   & $154 \pm 7$   & $8.9 \pm 0.7$         \\ 
$[$O\,{\sc i}] $\lambda 6364$   & $179 \pm 14$  & $3.3 \pm 0.4$         \\ 
$[$N\,{\sc ii}] $\lambda 6548$  & $165 \pm 11$  & $5.6 \pm 0.8$         \\ 
\Ha\ $\lambda 6563$             & $253 \pm 4$   & $46.7 \pm 1.2$        \\ 
$[$N\,{\sc ii}] $\lambda 6584$  & $204 \pm 7$   & $13.1 \pm 0.9$        \\ 
He\,{\sc i} $\lambda 6678$      & $177 \pm 38$  & $0.8 \pm 0.2 $        \\ 
$[$S\,{\sc ii}] $\lambda 6717$  & $113 \pm 40$  & $0.9 \pm 0.9$         \\ 
$[$S\,{\sc ii}] $\lambda 6731$  & $149 \pm 48$  & $0.9 \pm 0.4$         \\
\tableline
\multicolumn{3}{l}{\footnotesize{$^a$ in km s$^{-1}$}}                  \\
\multicolumn{3}{l}{\footnotesize{$^b$ in units of $10^{-16} \flux$}}    \\
\tableline 
\end{tabular}
\end{center}
\medskip

In addition to interstellar extinction, the observed C\,{\sc iv} and
Si\,{\sc iv} fluxes are further reduced by line absorption along the
line of sight to the supernova.  The absorption profiles for these
lines are known from {\it IUE\/} high-dispersion spectra of SN 1987A
taken at the time of outburst \citep{Welt99}. Since the intrinsic
profiles of the lines are poorly known, correcting for this absorption
is a somewhat uncertain endeavor. For now we just present the observed
fluxes and note that under reasonable assumptions of the intrinsic
profiles we find correction factors of $\approx 2$ for these lines
\citep{Pun00}.

\section{RADIATIVE SHOCK MODEL}

Figure~\ref{cartoon} illustrates the scenario that we believe accounts
for the emission from Spot 1.  The freely expanding supernova ejecta
are slowed from $\sim 15,000 \kms$ to $\sim 4,000 \kms$ by a reverse
shock.  The shocked gas drives a blast wave (forward shock), with
velocity $v_{\rm b} \approx 4,000 \kms$, into an H\,{\sc ii} region
having density $n_{\rm II} \approx 100 \pcc$
\citep{CD95,BBM97L,Lund99}.  The bright spots occur where the blast
wave encounters protrusions of the dense ($n \approx 10^4 \pcc$,
\citeauthor{LF96} \citeyear{LF96}) equatorial ring.  As it enters a
protrusion, the blast wave is slowed to a value $v_{\rm s} \approx
v_{\rm b} (n_{\rm II}/n)^{1/2} f(\theta)$, where $\theta$ is the angle
between the surface of the protrusion and the propagation direction of
the blast wave. The function $f(\theta)$, which accounts for shock
obliquity and the pressure increase due to the reflected shock, varies
from $f(0) \approx 2$ at the head of the protrusion to $f(\pi/2)
\approx 0.7$ along its sides \citep{BBM97}. We therefore expect a
range of shock velocities, $280 \,n_4^{-1/2} \kms \lae v_{\rm s} \lae
800 \,n_4^{-1/2} \kms$, to be present (where $n_4$ is the pre-shock
density in units of $10^4 \pcc$).

\medskip
\begin{center}
\includegraphics[width=3.25in]{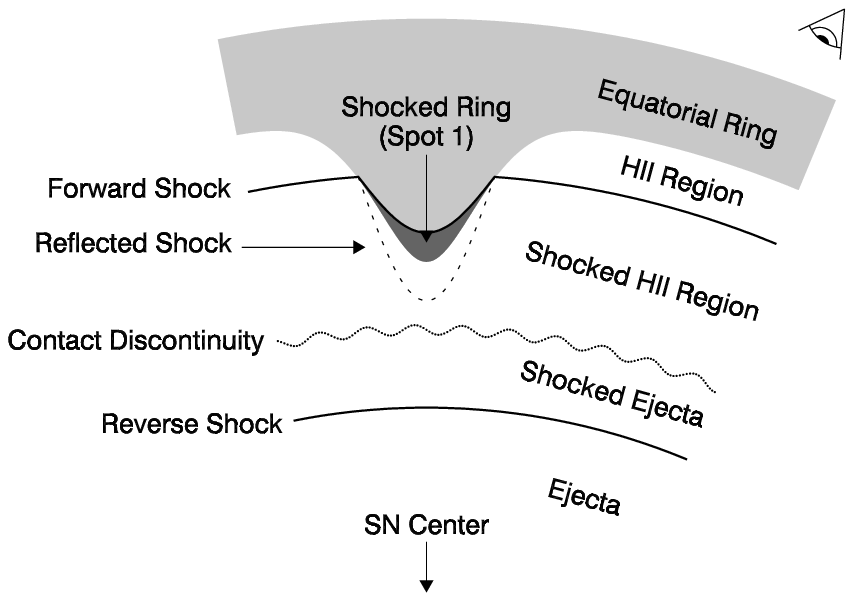}
\figcaption[f3.eps]{Cartoon of SNR 1987A's structure in the equatorial
plane.\label{cartoon}}
\end{center}
\medskip

An important parameter for the shocks considered here is the radiative
cooling time $t_{\rm c}$, defined as the time for the shocked gas to
cool from the post-shock temperature, $T_{\rm s} = 3\mu m_{\rm p}
v_{\rm s}^2/16k \approx 1.2 \times 10^6 (v_{\rm s} / 300 \kms)^2 \K$,
to $T_{\rm c} = 10^4 \K$.  A fit to the results from a plane-parallel
steady-state (1D-SS) shock code (e.g., \citeauthor{CR85}
\citeyear{CR85}) with abundances appropriate to SN 1987A
\citep{LF96,RD92} shows that, for $100 \kms < v_{\rm s} < 600 \kms$,
$t_{\rm c} \approx 4.6\, n_4^{-1} (v_{\rm s} / 300 \kms)^{3.7} \yr$.

When these spectra were taken, the age of the spot, $t_{\rm spot}
\approx 3 \yr$, was comparable to the cooling time for shocks with the
range of densities and velocities expected in the protrusion.  Shocks
with velocities $\gae 250\, n_4^{0.27} \kms$ have $t_{\rm c} > t_{\rm
spot}$ and would not yet have radiated away their energy, while slower
(radiative) shocks will have had time to cool and will convert all
their thermal energy into radiation.  Non-radiative shocks produce far
less ultraviolet and optical emission than radiative shocks.
Therefore, although faster shocks might be present, we expect the
ultraviolet and optical emission from Spot 1 to be dominated by shocks
with speeds $\lae 250\, n_4^{0.27} \kms$.

The radiative shock model does predict the presence of both strong UV
and optical lines. The ultraviolet lines produced in a radiative shock
come mainly from the cooling region, where the shocked gas's thermal
energy is radiated away.  The optical lines, however, are formed in
the cool layer of dense ($\sim n T_{\rm s}/T_{\rm c}$, where $T_c
\approx 10^{4} \K$) gas which forms behind the cooling region. This
photoabsorption layer quickly becomes thick enough to absorb the
downstream ionizing radiation produced in the cooling region and
converts the flux into optical emission lines.

The optical lines are formed with thermal widths characteristic of
$10^4 \K$ gas.  The actual profiles are much broader and must be due
to the macroscopic motion of fluid parcels in the photoabsorption
layer.  In a steady shock, the gas in the photoionization zone will
have the same velocity as the shock front, and therefore the optical
lines would be expected to have velocity profiles representative of
the hydrodynamics of the shock (specifically, those parts of the shock
that have developed a radiative layer).  The presence of both
redshifted and blueshifted emission can be explained by shocks
traveling both into and out of our line of sight (see
Figure~\ref{cartoon}).

\section{DISCUSSION}

It will not be an easy task to develop a quantitative model for the
spectrum and line profiles of the hot spot.  As Figure 3 illustrates,
these result from a superposition of shocks having a range of
velocities and aspect angles that depend on the unknown geometry and
density structure of the protrusion.  Our task is further complicated
by the fact that radiative shocks are subject to violent thermal and
dynamical instabilities (e.g., \citeauthor{IGF87} \citeyear{IGF87};
\citeauthor{KMC94} \citeyear{KMC94}; \citeauthor{WF96}
\citeyear{WF96}, \citeyear{WF98}), so steady shock models may be
inadequate to interpret the spectrum.

Despite the complexity of the actual situation, we can gain some
insight into the physical conditions in the spot by comparing the
actual spectra with the results from the 1D-SS shock code.  For
example, the code shows that the ratios of the flux of \NV\ to the
flux of lines from lower ionization states (\OIV, \NIV, and \CIV) are
sensitive functions of shock velocity.  N\,{\sc v} is not produced at
temperatures $T_{\rm s} \lae 2 \times 10^5 \K$ ($v_{\rm s} \lae 120
\kms$), above which its emissivity rises rapidly with shock
velocity. On the other hand, O\,{\sc iv}, N\,{\sc iv}, and C\,{\sc iv}
all form at lower temperatures and their emissivities are less
sensitive to velocity.  To reproduce the observed ratios with the
shock code we must constrain the shock velocity to the range $120 \kms
< v_{\rm s} < 150 \kms$.  But such a velocity range conflicts with the
observed optical line widths, which require radiative shocks with
velocities up to at least $\approx 250 \kms$.

However, only $\approx 20\%$ of the emission in the broad \Ha\ line
profile comes at high velocity ($|v| > 135 \kms$, measured from
line center). We may then argue that only $\approx 20\%$ of the \Ha\
emission must come from shocks faster than $135 \kms$.  If this is
true, the line ratios will be dominated by emission from the slower
shocks.  We find that if the slower shocks ($v_s < 135 \kms$) cover
about four times the surface area than the faster shocks ($v_s > 135
\kms$), then the net emission from these shocks produced ratios near
those observed.

The presence of both fast and slow radiative shocks indicates that the
protrusion must have a slightly higher density than the average ring
density.  Using hydrodynamical arguments presented in \S 3 we find
that preshock densities $n \gae 4 \times 10^{4} \pcc$ are required to
produce shocks as slow as $135 \kms$.  A less stringent lower limit is
provided by an independent argument based on the radiative cooling
time. Preshock densities of $n \gae 10^{4} \pcc$ are required for
shocks as fast as $250 \kms$ (as indicated by the \Ha\ line width) to
cool within $t_s \approx 3$ years.

We can also use the spectroscopic data to estimate the surface area of
the shock interaction.  Most of the emission comes from shocks with
velocities $\approx 135 \kms$, for which the shock code gives an \Ha\
surface emissivity of $0.4\, n_4 \flux$.  Correcting the observed \Ha\
flux for reddening \citep{Scud96} and assuming a distance of $50
\kpc$, we find that the shock must have an \Ha\ emitting surface area
of $A_s \approx 10^{33} \area$ (assuming $n = 4 \times 10^{4} \pcc$).
Then, assuming a hemispherical shape for the emitting surface, we
estimate that Spot 1 should have a characteristic dimension $D_{\rm
spot} \approx (2A_s/\pi)^{1/2} \approx 3 \times 10^{16} \cm$ (or
$\approx 0\farcs04$).  This is smaller than HST's resolution so
we are not surprised that the spot is spatially unresolved.

As Table 1 indicates, different emission lines have different widths.
This is expected, since the emissivities of the lines are sensitive
functions of shock velocities, and a range of shock velocities is
clearly in evidence.  Moreover, the differences among the optical line
widths may be explained by partial cooling in fast shocks. Emission
from ions like O\,{\sc i} and S\,{\sc ii} comes from regions downstream
of the gas that produces \Ha\ emission.  Faster shocks take longer to
develop these downstream regions.  The fastest shocks that produce
\Ha\ will not yet have developed their [O\,{\sc i}] and [S\,{\sc ii}]
emitting zones. Therefore, the [O\,{\sc i}] and [S\,{\sc ii}] lines
are narrower than the \Ha\ line.

Now that several new bright spots have appeared on the circumstellar ring
of SN 1987A \citep{LC00,GKC00,Lawr00}, we have new opportunities to
elucidate the development of radiative shocks.  By observing the
spectra of each of the spots, we can follow the development of shocks
having different ages and velocities.  The spots will continue to
brighten and their spectra will change.  The pressure that drives the
shocks will increase, the shocks will engulf more of the protrusions,
and faster shocks will develop radiative layers.  The most powerful
constraint on the hydrodynamics will be to resolve the line ratios as
a function of Doppler velocity. For example, the wings of the \NV\
line profile should be more prominent than those of \CIV.  As the
optical lines brighten, their wings should broaden.

For these reasons, it is imperative to observe the development of the
spectra of these spots on a regular basis.  These observations will
need to be performed by STIS in order to obtain UV line profiles
as well as spatially distinguish the spectra of the different spots.

\acknowledgements E.M. thanks S. Zhekov for fruitful
discussions. This research was supported by NASA through grants NAG
5-3313 and NTG 5-80 to the University of Colorado and grants GO-2563
and GO-7434 from the Space Telescope Science Institute, which is
operated by the Association of Universities for Research in Astronomy,
Inc., under NASA contract NAS 5-26555. C.S.J.P. acknowledges support
by the STIS IDT through NOAO by NASA.

\end{document}